\documentclass[prd,a4paper]{revtex4}
\usepackage{graphicx}
\usepackage{dcolumn}
\usepackage{amsmath}
\usepackage{bm}

\begin{document}

\title{Penrose-Carter diagram for an uniformly accelerated observer}

\author{Claude \surname{Semay}}
\thanks{FNRS Research Associate}
\email[E-mail: ]{claude.semay@umh.ac.be}
\affiliation{Groupe de Physique Nucl\'{e}aire Th\'{e}orique,
Universit\'{e} de Mons-Hainaut,
Acad\'{e}mie universitaire Wallonie-Bruxelles,
Place du Parc 20, BE-7000 Mons, Belgium}

\date{\today}

\begin{abstract}
An uniformly accelerated observer can build his proper system of
coordinates in a delimited sector of the flat Minkowski spacetime. The
properties of the position and time coordinate lines for such an
observer are studied and compared with the coordinate lines for an
inertial observer in a Penrose-Carter diagram for this spacetime.
\end{abstract}

\pacs{03.30.+p,04.20.Ha}
\keywords{Special relativity, Classical general relativity: Asymptotic
structure, Motion with a constant proper acceleration, Penrose-Carter
diagram}

\maketitle

\section{Introduction}
\label{sec:intro}

It is sometimes useful to dispose of a graphical representation of the
totality of the spacetime, for instance to study asymptotic forms of
various fields (metric, curvature tensor, electromagnetic field, etc.).
A very elegant mathematical technique to study the asymptotic properties
of spacetimes has been developed simultaneously by Roger Penrose and
Brandon Carter \cite{penr64,hawk73}. The idea is to perform what is
called a conformal transformation of spacetime to bring infinities at
finite distances while preserving its causal structure (light cones are
unaltered). Asymptotic calculations are then converted into calculations
at finite points with a set of new coordinates, the conformal
coordinates, which attribute finite values to infinities. Thereby, a
global picture of the causal structure for the totality of the studied
spacetime can be more easily obtained. The diagram of
the spacetime after transformation is called a Penrose-Carter diagram.

Students are sometimes confronted for the first time with Penrose-Carter
diagrams by studying the spacetime around a static or a rotating black
hole. Even if the flat Minkowski spacetime is studied with conformal
coordinates, exact coordinate line equations are often not given and
their properties are rarely studied. The purpose of this paper is to
perform a detailed study with conformal coordinates of the simplest
possible spacetime, the flat
Minkowski spacetime, from the point of view of an inertial observer and
an uniformly accelerated observer. The conformal coordinates and the
proper coordinates of the uniformly accelerated observer are both
obtained from a change of coordinates. But, these two transformations
have different physical contents which will discussed within the
text.

The intrinsic properties of a spacetime cannot depend on the
system of coordinates used to map this spacetime. But,
a better understanding of the structure of a spacetime can be obtained
by the knowledge of its coordinate lines with a clear physical meaning.
A natural choice is to take, when it is possible, lines with
constant time or position. For the flat Minkowski spacetime, these
coordinate lines are straight lines
cutting each other at right angle in an ordinary diagram while, in a
Penrose-Carter diagram, the equations of these lines are complicated
functions of the conformal coordinates. These are studied in
section~\ref{sec:mink}.

The motion of an observer with a constant proper acceleration can be
treated analytically. It is a classical exercise of special
relativity that can be found in many textbooks
\cite{sear68,misn73,rind77,sema05}. In this framework, the very
prominent notion of event horizon can be introduced in a simpler context
than the one of black hole for instance.
An uniformly accelerated observer can build his proper system of
coordinates in a delimited, but infinite, sector of the flat Minkowski
spacetime \cite{rind66,desl87,desl89,frol98,sema06}. 
The corresponding time and
position coordinate lines are respectively hyperbolas and straight lines
in an ordinary diagram. A detailed study of the equations of these lines
with conformal coordinates is performed in section~\ref{sec:uao}.

A brief summary of our results about these mappings of the flat
Minkowski spacetime for both an inertial observer and an uniformly
accelerated observer is given in section~\ref{sec:sum}.

Penrose-Carter diagrams are generally used with spatial spherical
coordinates plus a time coordinate. In this case, the study of only
radial trajectories of particles, with angular coordinates fixed, is
performed in a $1+1$ Minkowski spacetime. In this paper, we will
consider rectangular coordinates plus a time coordinate. As we will
look only at motions along the $x$-axis, keeping $y$ and $z$ coordinates
fixed, all results are also presented in a $1+1$ Minkowski spacetime.

\section{The Minkowski spacetime}
\label{sec:mink}

\subsection{Change of coordinates}
\label{sec:chgcoo}

The usual change of coordinates to bring back infinities at finite
distances is
\begin{eqnarray}
\label{psixi1}
ct + x &=& L \tan(\psi+\xi), \nonumber \\
ct - x &=& L \tan(\psi-\xi),
\end{eqnarray}
with $-\pi/2 < \psi \pm \xi < \pi/2$. $L$ is an arbitrary length, and
$\psi$ and $\xi$ are
dimensionless quantities, the conformal coordinates. It is
useful to define new dimensionless spacetime variables: $T=ct/L$ and
$X=x/L$ which will be used throughout the text.
Equations~(\ref{psixi1}) are
then reduced to
\begin{eqnarray}
\label{psixi2}
T + X &=& \tan(\psi+\xi), \nonumber \\
T - X &=& \tan(\psi-\xi).
\end{eqnarray}
With these conformal coordinates, the totality of the spacetime is
represented by a square (see figure~\ref{fig:uao1}), sometimes called
the ``Minkowski diamond". As we will see below, coordinate lines with
constant position and time converge at the apexes of the square. These
points are the conformal infinities for space and time.

The equation of motion of a photon passing by the position $X_*$ at time
$T_*$ is $X\pm T=X_*\pm T_*$. In the conformal coordinates, it is
written
\begin{equation}
\label{wllight}
\xi\pm \psi= \arctan(X_*\pm T_*).
\end{equation}
So, as in an ordinary diagram, the world line of a photon in the
Penrose-Carter
diagram is still a line slanted at an angle of 45$^\circ$, but with
respect to the
$\xi$ and $\psi$ coordinates. The causal structure in the Penrose-Carter
diagram is given by light cones for variables $\xi$ and $\psi$, as it is
given by light cones for variables $X$ and $T$ in an ordinary diagram.
The diagonal boundaries of the Penrose-Carter diagram are the infinities
where world line of light rays must end.

With the new coordinates, the metric $ds^2=c^2dt^2-dx^2$ is written
\begin{equation}
\label{ds2}
ds^2=L^2(dT^2-dX^2)=L^2\frac{d\psi^2-d\xi^2}{\cos^2(\psi+\xi)\cos^2(\psi
-\xi)}.
\end{equation}

\subsection{Coordinate lines}
\label{sec:cool}

\begin{center}
\begin{figure}[tbh]
\includegraphics*[height=8cm]{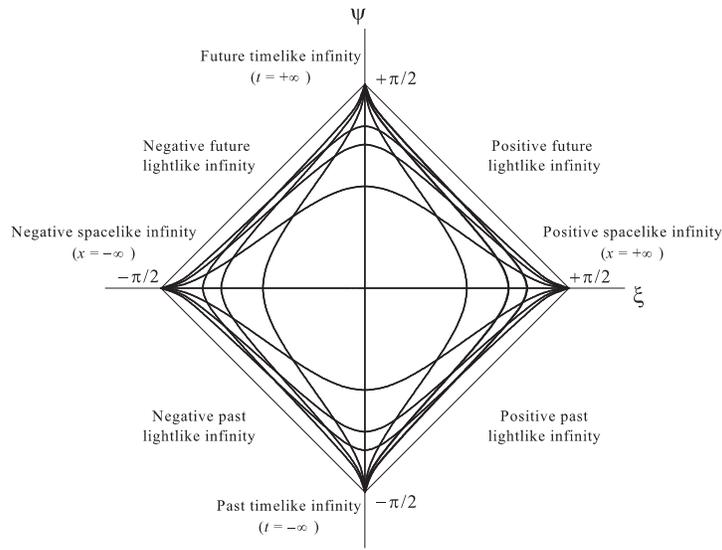}
\caption{Coordinate lines for the Minkowski spacetime in the
Penrose-Carter
diagram. From left to right, lines for constant positions
$X_*=-3, -2, -1, 0, 1, 2, 3$ are indicated. From bottom to top, lines
for constant times $T_*=-3, -2, -1, 0, 1, 2, 3$ are indicated. The
$X_*=0$ ($T_*=0$) line is the $\xi$=0 ($\psi=0$) line.
\label{fig:uao1}}
\end{figure}
\end{center}

Following equations~(\ref{psixi2}), a time coordinate line with the
constant position $X_*$ is given by
\begin{equation}
\label{xstar}
2 X_* = \tan(\psi+\xi) - \tan(\psi-\xi).
\end{equation}
Using the properties of the tangent function, this relation can be
recast into the form
\begin{equation}
\label{xiflat}
\xi(\psi,X_*)=\arctan\left[ \frac{-(1+\tan^2\psi) + \sqrt{(1+\tan^2\psi)
^2+4 X_*^2 \tan^2\psi}}{2 X_* \tan^2\psi} \right],
\end{equation}
with $\psi \in\ ]-\pi/2,\pi/2[$ and $X_* \in\ ]-\infty,\infty[$ (see
figure~\ref{fig:uao1}). This
function is even for the variable $\psi$ and odd for the variable $X_*$.
It is vanishing at the conformal time infinities,
$\xi(\pm \pi/2,X_*)=0$, and $\xi(0,X_*)=\arctan X_*$.

The derivative of function~(\ref{xiflat}) with respect to $\psi$ is
written
\begin{equation}
\label{xiflatp}
\partial_\psi \xi(\psi,X_*)=\frac{-\sqrt{2}X_* \sin(2\psi)}
{\sqrt{2+X_*^2(1-\cos(4\psi))}}.
\end{equation}
It has three zeros for all values of $X_*$:
$\partial_\psi \xi(\pm \pi/2,X_*)=\partial_\psi \xi(0,X_*)=0$.
The slopes at extremities of these coordinate lines are then vanishing.

For infinite values of $X_*$, we obtain
\begin{equation}
\label{xiflatlim}
\lim_{X_*\to\pm\infty} \xi(\psi,X_*) =
\pm \arctan \left( | \cot \psi | \right),
\end{equation}
that is to say, $\xi=\pm (\psi+\pi/2)$ for $\psi \in [-\pi/2,0]$ and
$\xi=\pm (\pi/2-\psi)$ for $\psi \in [0,\pi/2]$. These are the
boundaries of the spacetime, as expected. Let us note that we have
\begin{equation}
\label{xiflatplim}
\lim_{X_*\to\pm\infty} \partial_\psi \xi(\psi,X_*) =
\left\{ \
\begin{array}{l}
\phantom{-}0 \ \textrm{for} \ \psi=-\pi/2,0,\pi/2\\
\pm 1\ \textrm{for} \ \psi\in\ ]-\pi/2,0[ \\
\mp 1\ \textrm{for} \ \psi\in\ ]0,\pi/2[
\end{array}\right. ,
\end{equation}
in agreement with the results given just above.

A space coordinate line with the constant
time $T_*$ is given by\begin{equation}
\label{tstar}
2 T_* = \tan(\psi+\xi) + \tan(\psi-\xi).
\end{equation}
The transformations $T_* \leftrightarrow X_*$ and
$\psi \leftrightarrow \xi$ change this equation into
equation~(\ref{xstar}).
So there is a complete symmetry between time and space coordinate lines,
as expected.
Their properties are the same and the discussion above can be completely
adapted to the space coordinate lines.

\section{The uniformly accelerated observer}
\label{sec:uao}

\subsection{Hyperbolic motion}
\label{sec:hypm}

Let us consider an uniformly accelerated observer with a constant proper
acceleration with magnitude $A>0$. Its motion is such that it reaches
the point $x=0$ in the inertial frame at time $t=0$ with a vanishing
speed. The world line of the observer is given by
\cite{sear68,misn73,rind77,sema05,rind66,desl87,desl89,frol98,sema06}
\begin{equation}
\label{hypm}
\left(\frac{A\,x}{c^2} + 1\right)^2 - \left(\frac{A\,ct}{c^2}\right)^2
=1,
\end{equation}
which is the equation of a branch of hyperbola in spacetime. So this
motion is also called hyperbolic.

\begin{center}
\begin{figure}[tbh]
\includegraphics*[height=8cm]{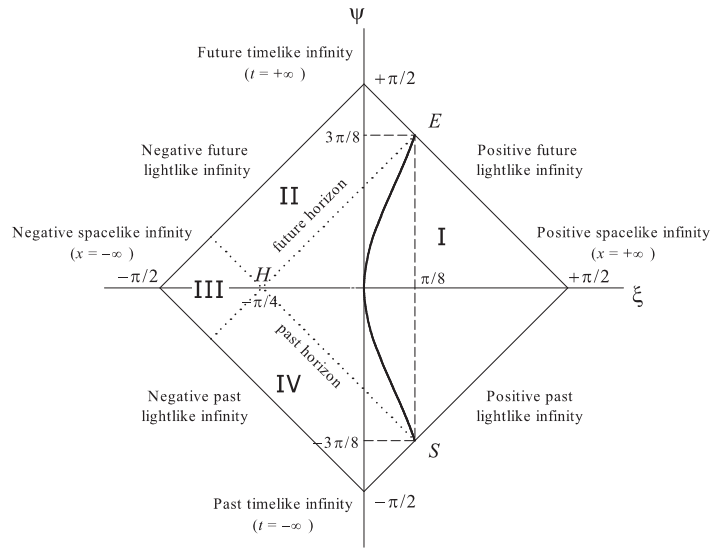}
\caption{World line of the uniformly accelerated observer in the
Penrose-Carter
diagram for the Minkowski spacetime. Past and future event horizons are
indicated with the four Rindler sectors. \label{fig:uao2}}
\end{figure}
\end{center}

If we choose $L=c^2/A$, this equation can be recast into the form
\begin{equation}
\label{hypmred}
\left(X + 1\right)^2 - T^2 =1.
\end{equation}
The asymptotes of this curve are the two straight lines with equations
$T=\pm (X +1)$. Consequently, the asymptotes of the world line of the
accelerated observer defines two event horizons
\cite{sear68,misn73,rind77,sema05,rind66,desl87,desl89,frol98,sema06}. 
In the conformal coordinates, equations of these horizons are
\begin{equation}
\label{horiz}
\psi = \pm (\xi+\pi/4).
\end{equation}
These two horizons cross at event $H$ with spacetime coordinates
$(\xi_H,\psi_H)=(0,-\pi/4)$ corresponding to $(X,T)=(-1,0)$. The past
(future) horizon intercepts the positive past (future) light infinity at
event $S$ ($E$) with spacetime coordinates
$(\xi_S,\psi_S)=(\pi/8,-3\pi/8)$ ($(\xi_E,\psi_E)=(\pi/8,3\pi/8)$). They
cut the whole spacetime in four regions, called Rindler sectors (see
figure~\ref{fig:uao2}) \cite{frol98}. The sector~I is 
the portion of the spacetime in
which the uniformly accelerated observer lives: he can send information
to any event and he can receive information from any event in this
sector. In sector~II, located ``above" the future horizon, the observer
can send information to any event but cannot receive information from
this sector. The situation is exactly the opposite in the sector~IV,
located ``below" the past horizon. The sector~III is causally completely
disconnected from the observer.

Written in the conformal coordinates, equation~(\ref{hypmred}) becomes
\begin{equation}
\label{hypmpen}
- \tan(\psi+\xi) \tan(\psi-\xi) + \tan(\psi+\xi) - \tan(\psi-\xi)=0.
\end{equation}
This relation can be recast into the form
\begin{equation}
\label{hypmpen2}
\xi=\arctan\left[ -(1+\tan^2\psi) + \sqrt{(1+\tan^2\psi)
^2+ \tan^2\psi} \right].
\end{equation}
One can check that this world line starts at the event $S$ and ends at
the event $E$, both on positive lightlike infinity, as expected since
the speed of this observer is equal to the speed of light in the
infinite past and future.

The uniformly accelerated observer can build his proper system of
dimensionless coordinates, spacelike $X_0$ and timelike $T_0$, valid
only in sector~I \cite{rind66,desl87,desl89,frol98,sema06}. 
The change of
coordinates between $(T,X)$ and $(T_0,X_0)$ is given by \cite{sema06}
\begin{eqnarray}
T=(X_0 +1) \sinh T_0, \nonumber  \\
\label{xtx0t01}
X+1= (X_0 +1) \cosh T_0.
\end{eqnarray}
The corresponding metric is
\begin{equation}
\label{ds2uao}
ds^2=L^2 \left[ (X_0 +1)^2 dT_0^2 - dX_0^2 \right].
\end{equation}
With relations~(\ref{xtx0t01}), it is possible to
determine, in the inertial frame, the equations of the coordinate
lines of this observer proper frame. In this last
frame, the equation of a time coordinate line with $X_0$ constant is
\begin{equation}
\label{lctemps}
\left( X + 1 \right)^2 - T^2 = \left( X_0 + 1 \right)^2.
\end{equation}
This curve is a branch of hyperbola whose asymptotes are the two event
horizons mentioned above. These horizons are located on the degenerate
asymptotes obtained with $X_0=-1$ in equation~(\ref{lctemps}).
Obviously, the world line of the uniformly accelerated observer in his
proper frame is given by $X_0=0$. Let us note that an object with a
constant position $X_0$ is not at rest with the uniformly accelerated
observer. This object is characterized by a proper uniform acceleration
whose magnitude $A(X_0)$ is given by
\cite{desl89,sema06}
\begin{equation}
\label{ax0}
A(X_0) = \frac{A}{1+X_0}.
\end{equation}
In the same way, the distance between the uniformly accelerated observer
and an object with the same proper acceleration $A$ varies exponentially
with the proper time of the observer \cite{sema06}.

\begin{center}
\begin{figure}[tbh]
\includegraphics*[height=8cm]{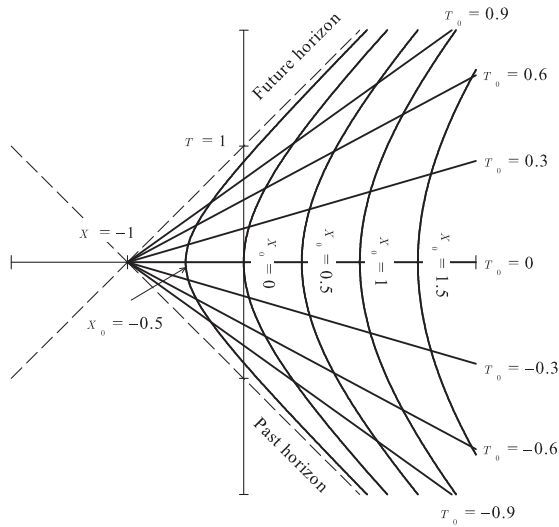}
\caption{Coordinate lines, associated with an uniformly accelerated
observer, for constant times $T_0$ (straight lines) and for constant
positions $X_0$ (hyperbolas), in an inertial frame. The world line of
the observer is the coordinate line $X_0=0$. \label{fig:uao3}}
\end{figure}
\end{center}

The equation of a position coordinate line with $T_0$ constant is
\begin{equation}
\label{lcxpos}
T = \left( X + 1 \right) \tanh T_0 .
\end{equation}
This is a straight line containing the event $(X,T)=(-1,0)$, that is to
say the event $H$, the intersection of the two event horizons. Some
coordinate lines are drawn in figure~\ref{fig:uao3}. It can be seen that
the future horizon and the past horizon correspond respectively to the
position coordinate lines $T_0=+\infty$ and $T_0=-\infty$. Both horizons
form also the space coordinate line $X_0=-1$.

Transformations~(\ref{psixi2}) and (\ref{xtx0t01}) are both changes of
coordinates from the usual coordinates $(T,X)$ defined in the inertial
frame of the flat Minkowski spacetime. But, their physical content is
strongly different.

With the change of coordinates~(\ref{psixi2}), bringing back infinities
at finite distances, the coordinates lines are heavily distorted but
the coordinates $(\psi,\xi)$ are just another set of coordinates for the
inertial observer in the whole flat Minkowski spacetime,
with the particularity that light cones are preserved. To some extent,
the distortions produced in the transformation
$(T,X)\rightarrow (\psi,\xi)$ are like those obtained for the
transformation in the plane passing from Cartesian coordinates $(x,y)$
to polar coordinates $(r,\theta)$: A straight line $y=a x+b$ in the
plane is generally not represented by a straight line in a $(r,\theta)$
diagram.

The coordinates $(T_0,X_0)$, coming from the change of
coordinates~(\ref{xtx0t01}), are the proper coordinates in the
non inertial frame of an uniformly accelerated observer and are only
valid in a limited sector of spacetime. For this observer, all objects
in free motion in the Minkowski spacetime undergo an accelerated motion
and the light ray world lines are given by exponential functions
\cite{desl87}. The uniformly accelerated observer has the impression
that an uniform gravitational field exists in his surroundings, although
the spacetime is in reality always flat \cite{desl89}.

Now, we will express the coordinate lines~(\ref{lctemps}) and
(\ref{lcxpos}) with the conformal coordinates in order to draw them in
the Penrose-Carter diagram. In the calculations, the following two
identities will be useful:
\begin{equation}
\label{tanpi8}
\tan\frac{\pi}{8}=\sqrt{2}-1 \quad \textrm{and} \quad
\tan\frac{3\pi}{8}=\sqrt{2}+1.
\end{equation}

\subsection{Position coordinate lines}
\label{sec:pos}

\begin{center}
\begin{figure}[tbh]
\includegraphics*[height=8cm]{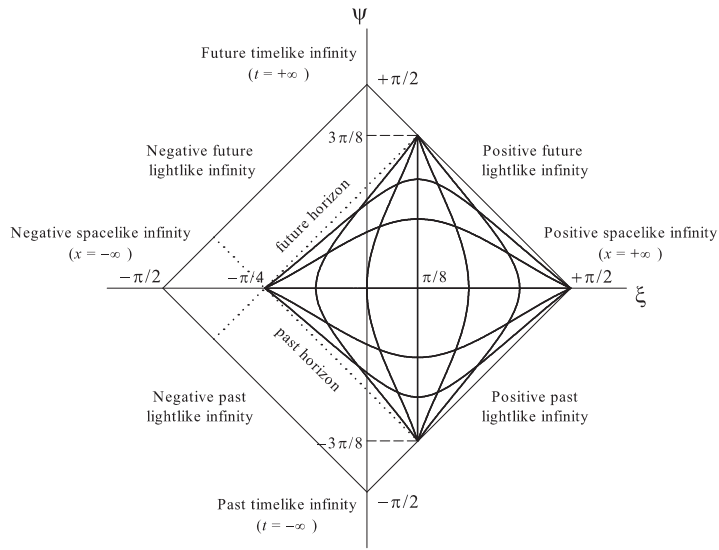}
\caption{Coordinate lines associated with the uniformly accelerated
observer, for constant times $T_0$ and for constant positions $X_0$, in
the Penrose-Carter diagram for the Minkowski spacetime. From bottom to
top, constant time lines for $T_0=-1, -0.5, 0, 0.5, 1$ are indicated.
From left to right, constant position lines for
$X_0=\tan \frac{s\pi}{8}$ with $s=-1,0,1,2,3$ are
indicated. The $T_0=0$ line is the $\psi=0$ line. The world line of this
observer is the coordinate line $X_0=0$. \label{fig:uao4}}
\end{figure}
\end{center}

With the conformal coordinates, equation~(\ref{lcxpos}) is given by
\begin{equation}
\label{pcl1}
\tan(\psi+\xi) + \tan(\psi-\xi) = \tanh T_0 \left[ \tan(\psi+\xi) - \tan
(\psi-\xi) +2 \right].
\end{equation}
After some calculations, the position coordinate line equation, with
$T_0$ constant, for the uniformly accelerated observer (uao) can be
written
\begin{equation}
\label{pcl2}
\psi_{\textrm{uao}}(\xi,T_0)=\arctan\left[ \frac{1+\tan^2\xi - \sqrt
{(1+\tan^2\xi)^2-4 \tanh^2 T_0 \tan \xi (1-\tan^2\xi)}}{2 \tanh T_0 \tan
\xi (1- \tan\xi)} \right],
\end{equation}
with $\xi \in\ ]-\pi/4,\pi/2[$ and $T_0\in\ ]-\infty,\infty[$ (see
figure~\ref{fig:uao4}). This function is vanishing at the two spacelike
extremities of sector~I,
$\psi_{\textrm{uao}}(-\pi/4,T_0)=\psi_{\textrm{uao}}(\pi/2,T_0)=0$. It
is odd for the variable $T_0$ and, obviously, we have
$\psi_{\textrm{uao}}(\xi,0)=0$.
The position coordinate line with $T_0=0$ cuts sector~I in two equal
parts. We can expect that $\psi_{\textrm{uao}}$
is even for the
variable $\xi$ with respect to $\pi/8$, the middle of the interval
$]-\pi/4,\pi/2[$. If we define the new variable $y=\xi-\pi/8$, it can
be checked, after a tedious calculation, that
$\psi_{\textrm{uao}}(y,T_0)=\psi_{\textrm{uao}}(-y,T_0)$.
We can compute that
$\psi_{\textrm{uao}}(\pi/8,T_0)=\arctan\left[ (1+\sqrt{2}) \tanh
\frac{T_0}{2} \right]$, which implies that
$\lim_{T_0\to \pm\infty}\psi_{\textrm{uao}}(\pi/8,T_0)=\pm 3\pi/8$; the
two timelike infinities of sector~I are reached.

The derivative of function~(\ref{pcl2}) with respect to $\xi$ is given
by
\begin{equation}
\label{derivpcl}
\partial_\xi\psi_{\textrm{uao}}(\xi,T_0)=\frac{\cos^2\xi \tanh T_0
\left(1-\tan\xi(2+\tan \xi) \right)}{\sqrt{1-\sin(4 \xi)\tanh^2 T_0}}.
\end{equation}
Due to the symmetry properties of the position coordinate lines, we have
$\partial_\xi\psi_{\textrm{uao}}(\pi/8,T_0)=0$. But, at the two
spacelike extremities of sector~I, the derivatives are not vanishing and
varies between $-1$ and $1$:
$\partial_\xi\psi_{\textrm{uao}}(-\pi/4,T_0)=\tanh T_0$ and
$\partial_\xi\psi_{\textrm{uao}}(\pi/2,T_0)=-\tanh T_0$. When
$T_0\to \pm\infty$, the position coordinate lines form the edges of
sector~I.

\subsection{Time coordinate lines}
\label{sec:tim}

With the conformal coordinates, equation~(\ref{lctemps}) is given by
\begin{equation}
\label{tcl1}
-\tan(\psi+\xi) \tan(\psi-\xi) + \tan(\psi+\xi) - \tan
(\psi-\xi) = Y_0,
\end{equation}
with $Y_0=X_0(X_0+2)$.
After some calculations, the time coordinate line equation, with
$X_0$ constant, can be written
\begin{equation}
\label{tcl2}
\xi_{\textrm{uao}}(\psi,X_0)=\arctan\left[ \frac{-(1+\tan^2\psi) + \sqrt
{(1+\tan^2\psi)^2 + (1+ Y_0\tan^2 \psi) (Y_0+\tan^2\psi)}}
{1+ Y_0 \tan^2\psi} \right].
\end{equation}
with $\psi \in\ ]-3\pi/8,3\pi/8[$ and $X_0 \in\ ]-1,\infty[$ (see
figure~\ref{fig:uao4}). This function is even for the variable $\psi$.
Since $\xi_{\textrm{uao}}(0,X_0)=\arctan X_0$, we have
$\lim_{X_0\to -1}\xi_{\textrm{uao}}(0,X_0)=-\pi/4$ and
$\lim_{X_0\to \infty}\xi_{\textrm{uao}}(0,X_0)=\pi/2$; the two spacelike
infinities of sector~I are reached. Because
$\xi_{\textrm{uao}}(\pm3\pi/8,X_0)=\pi/8$,
the coordinate lines~(\ref{tcl2}) connect the two events $S$ and $E$,
the timelike infinities of sector~I.

The derivative of function~(\ref{tcl2}) with respect to $\xi$ is given
by
\begin{equation}
\label{derivtcl}
\partial_\psi \xi_{\textrm{uao}}(\psi,X_0)=\frac{\left( 1-Y_0 \right)
\tan \psi}{\sqrt{\sec^4\psi+(1+ Y_0\tan^2 \psi) (Y_0+\tan^2\psi)}}.
\end{equation}
As expected from the symmetry properties of the position coordinate
lines, we have $\partial_\psi \xi_{\textrm{uao}}(0,T_0)=0$.
But, at the two
timelike extremities of sector~I, the derivatives are not vanishing and
varies between $-1$ and $1$:
\begin{equation}
\label{pdef1}
\partial_\psi \xi_{\textrm{uao}}(\pm 3\pi/8,X_0)=\pm p(X_0)
\quad \textrm{with}\quad p(X_0)=\frac{1-X_0(X_0+2)}{3+X_0(X_0+2)}.
\end{equation}
We have $\lim_{X_0\to -1}p(X_0)=1$ and $\lim_{X_0\to \infty}p(X_0)=-1$.
When the position $X_0$ reaches its extremal values, the time coordinate
lines form also the edges of sector~I.

We can also remark that $p(\tan \frac{\pi}{8})=0$, and it can be shown
that $\xi_{\textrm{uao}}(\psi,\tan \frac{\pi}{8})=\pi/8$. We can
wonder if $\xi_{\textrm{uao}}$ is odd for the variable $X_0$ with
respect
to $\tan \frac{\pi}{8}$. If we define the new variable $u$ with
$X_0=\tan(\frac{\pi}{8}-u)$ and the new function $w$ by
$\xi_{\textrm{uao}}=\pi/8-w$, it can be checked, after a tedious
calculation, that $w(u)$ is an odd function of $u$. The time coordinate
line with $X_0=\tan \frac{\pi}{8}$ cuts sector~I in two equal parts.

\subsection{Link between coordinate lines}
\label{sec:link}

By looking at figure~\ref{fig:uao4}, it seems that the position and time
line coordinates are very similar. So we can study the differences
between the functions $\xi_{\textrm{uao}}(a,X_0)-\pi/8$ and
$\psi_{\textrm{uao}}\left(a+\pi/8,T_0\right)$ with
$a \in\ ]-3\pi/8,3\pi/8[$. Suitable translations are made in order that
both functions coincide at their extremities: $a = \pm 3\pi/8$. To
perform a
comparison, a link must be found between variables $X_0$ and $T_0$.
Knowing the domain of each of these quantities, we can try
\begin{equation}
\label{t0x0}
T_0(X_0)=\arg\tanh \left[ \frac{8}{3\pi} \left( \arctan
X_0-\frac{\pi}{8} \right) \right].
\end{equation}
We have then $T_0(X_0=-1)=-\infty$, $T_0(X_0=\tan \frac{\pi}{8})=0$ and
$T_0(X_0=\infty)=\infty$.

\begin{center}
\begin{figure}[tbh]
\includegraphics*[height=4cm]{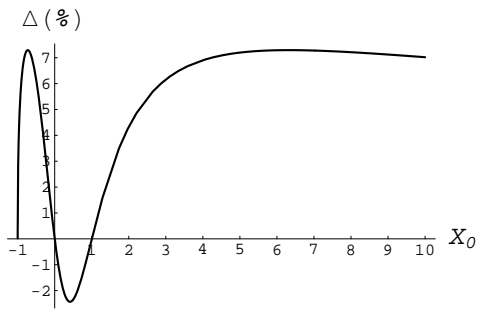}
\caption{Maximal relative gap $\Delta$ (see formula~(\ref{delta}))
as a function of $X_0$. \label{fig:uao5}}
\end{figure}
\end{center}

Let us define the two new functions
\begin{eqnarray}
\label{f1f2a}
f_T(a,b)&=& \xi_{\textrm{uao}}(a,b)-\frac{\pi}{8},   \\
\label{f1f2b}
f_X(a,b)&=& \psi_{\textrm{uao}}\left(a+\frac{\pi}{8},T_0(b)\right),
\end{eqnarray}
with $a \in\ ]-3\pi/8,3\pi/8[$ and $b\in\ ]-1,\infty[$. Thanks to
equation~(\ref{t0x0}),
these functions are built in such a way
that $f_T(a,b)-f_X(a,b)$ is vanishing for all values of $a$
when $b=-1$, $\tan \frac{\pi}{8}$ and $\infty$, that is to say when the
corresponding coordinate lines
form the borders of sector~I and when they cut
symmetrically this sector. As
$f_T(\pm 3\pi/8,b)=f_X(\pm 3\pi/8,b)=0$ and as these
functions are even
in $a$, we define the maximal relative gap
$\Delta$ between $f_T$ and $f_X$ by the formula
\begin{equation}
\label{delta}
\Delta(b) = 2 \frac{f_T(0,b)-f_X(0,b)}{f_T(0,b)+f_X(0,b)}.
\end{equation}
We can see on figure~\ref{fig:uao5} that the gap is always small. We
could
expect that $\Delta(\tan \frac{\pi}{8}) = 0$, but this not the case
because
$f_T(0,\tan \frac{\pi}{8})=f_X(0,\tan \frac{\pi}{8})=0$. We can also
remark that
$\Delta(0)=\Delta(1)=0$. Actually, it is possible to show, after some
lengthy calculations, that $f_T(a,b)=f_X(a,b)$ when $b=0$ and 1, for all
values of $a$. Finally, we have
$\xi_{\textrm{uao}}(a,X_0)-\pi/8=\psi_{\textrm{uao}}
(a+\pi/8,T_0)$ for $(X_0,T_0)=(-1,-\infty)$,
$(0,-\arg\tanh \frac{1}{3})$, $(\tan \frac{\pi}{8},0)$,
$(1,\arg\tanh \frac{1}{3})$ and $(\infty,\infty)$.

\section{Summary}
\label{sec:sum}

In this paper, a detailed study of the simplest possible spacetime, the
flat Minkowski spacetime, is performed with conformal coordinates, from
the point of view of an inertial observer and an uniformly accelerated
observer. Equations for coordinate lines with constant time or position
are given and their properties are studied.

Transformations~(\ref{psixi2}) and (\ref{xtx0t01}) are both possible
changes of coordinates in an inertial frame of the flat Minkowski
spacetime. Conformal coordinates from eqs.~(\ref{psixi2}) are just
another set of coordinates for an inertial observer but they allow a
clear interpretation of the causal structure for the whole spacetime.
The coordinates from eqs.~(\ref{xtx0t01}) are the proper coordinates of
an uniformly accelerated observer in a limited sector of the flat
Minkowski spacetime; They allow to study the motion of particles as seen
by this observer.

The Penrose-Carter diagram of the flat Minkowski spacetime for an
inertial
observer looks like a diamond, in which coordinate lines connect
opposite apexes which are the spacelike and timelike infinities.
Borders of this diamond are the lightlike infinities where world lines
of light rays end. The proper spacetime of an uniformly accelerated
observer is a small diamond included in the first one with one common
spacelike infinity. At first sight, the coordinate lines for such an
observer seems similar to those for the whole spacetime (compare
figure~\ref{fig:uao1} with figure~\ref{fig:uao4}) but there are big
differences:
\begin{itemize}
\item The time coordinate lines for the uniformly accelerated
observer end on the lightlike infinities of the whole spacetime,
while position coordinate lines end on one extremity at a spacelike
infinity and on the other extremity, due to the horizons, at a finite
position.
\item Considered as functions of $\xi$ or $\psi$, the slope at
extremities of the coordinate lines for the uniformly
accelerated observer varies from $-1$ to $1$, while the slope at
extremities is vanishing for the coordinate lines of
the inertial observer.
\item There is not a perfect symmetry between time and position
coordinate lines for the uniformly accelerated observer as it is the
case for the coordinate lines of the inertial observer.
\end{itemize}
All these differences could be expected from the examination of
figure~\ref{fig:uao3}. But, in this paper, the equations and properties
of the coordinate lines for both an inertial observer and an uniformly
accelerated observer are given for the flat Minkowski spacetime in a
Penrose-Carter diagram. This can help to understand the beautiful
properties of the conformal transformation associated with this kind of
diagram.


\end{document}